\documentclass[runningheads]{llncs}
\usepackage{bbm}
\usepackage{xspace}
\usepackage{stmaryrd}
\usepackage{graphicx}
\usepackage{hyperref}
\usepackage{booktabs}
\usepackage[dvipsnames]{xcolor}
\usepackage{array, multirow, boldline}
\usepackage{caption}
\usepackage{subcaption}
\usepackage{todonotes}
\usepackage{float}
\restylefloat{table}

\newcommand{\anon}[1]{\textsc{anonymized}\xspace}

\hypersetup{
    colorlinks = true,
    linkcolor=red,
    citecolor=red,
    urlcolor=blue,
    linktocpage 
}
\begin{document}

%
\title{Numerical Uncertainty of Convolutional Neural Networks Inference for Structural Brain MRI Analysis}
\titlerunning{Numerical Uncertainty of Convolutional Neural Networks}
%
%
\author{Inés Gonzalez Pepe \and
Vinuyan Sivakolunthu \and
Hae Lang Park \and\\
Yohan Chatelain \and
Tristan Glatard}
\authorrunning{I. Gonzalez Pepe et al.}
%
\institute{Department of Computer Science and Software Engineering,\\ Concordia University, Montreal, Canada }
%
\maketitle              
\begin{abstract}
  This paper investigates the numerical uncertainty of Convolutional Neural Networks (CNNs) inference for structural brain MRI analysis. It applies Random Rounding---a stochastic arithmetic technique--- to CNN models employed in non-linear registration (SynthMorph) and whole-brain segmentation (FastSurfer), and compares the resulting numerical uncertainty to the one measured in a reference image-processing pipeline (FreeSurfer recon-all). Results obtained on 32  representative subjects show that CNN predictions are substantially more accurate numerically than traditional image-processing results (non-linear registration: 19 vs 13 significant bits on average; whole-brain segmentation: 0.99 vs 0.92  Sørensen-Dice score on average), 
  which suggests a better reproducibility of CNN results across execution environments. 

\end{abstract}

\section{Introduction}
A motivating factor to study numerical uncertainty in neuroimaging is to establish measures of reliability in the tools observed, particularly in light of the reproducibility crisis \cite{botvinik2020variability,Fanelli_2017,baker_2016}.
Numerical uncertainty is key to the robustness of
neuroimaging analyses. Small computational perturbations introduced in execution environments--- including operating systems,
hardware architecture, and parallelization---may amplify throughout analytical pipelines and result
in substantial differences in the final outcome of analyses~\cite{gronenschild2012effects,glatard2015reproducibility}. Such
instabilities have been observed across many different tools and imaging modalities~\cite{salari2021accurate,kiar2021numerical}, and are likely to impact the reproducibility and robustness of analyses.

Convolutional Neural Networks (CNNs) are increasingly adopted for  registration~\cite{hoffmann2021synthmorph,iglesias2023ready,balakrishnan2019tmi} and segmentation~\cite{roy2019quicknat,henschel2020fastsurfer,jog2019psacnn,li2017compactness} of structural MRIs. 
Once trained, CNNs are orders of magnitude faster than traditional image-processing methods,
achieve comparable accuracy, and seem to exhibit better generalizability to image modalities and orientations. However, the numerical uncertainty associated with CNN predictions in neuroimaging remains largely unexplored. While previous works suggested that CNNs might be subject to numerical instability~\cite{higham2002accuracy,kloberdanz2022deepstability,chakraborty2021survey}, it is unclear how such instabilities manifest in specific CNN architectures used in structural brain MRI, and how the resulting numerical uncertainty compares to the one of traditional methods.

This paper measures the numerical uncertainty associated with CNN inference in neuroimaging, focusing specifically on non-linear registration and whole-brain segmentation of structural MRIs. To do so, it applies Random Rounding (RR)~\cite{forsythe1959reprint,fevotte2016verrou}---a practical stochastic arithmetic technique to estimate numerical uncertainty---to state-of-the-art CNN models SynthMorph~\cite{hoffmann2021synthmorph} and FastSurfer~\cite{henschel2020fastsurfer}, and compare their numerical uncertainty to the one measured from the FreeSurfer~\cite{fischl2002whole} ``recon-all" reference neuroimaging tool.

\section{Materials \& Methods}

We measured the numerical uncertainty of CNN models SynthMorph (non-linear registration) and FastSurfer (whole-brain segmentation) using RR. We applied these models to 35 subjects randomly selected in the CoRR dataset, using the FreeSurfer recon-all pipeline as a baseline for numerical uncertainty comparison. 

\subsection{Random Rounding}


Random Rounding (RR)~\cite{forsythe1959reprint} is a form of Monte-Carlo Arithmetic (MCA)~\cite{parker1997monte} that simulates rounding errors by applying the following perturbation to all floating-point (FP) operations of an application:
\[
  random\_rounding(x \circ y) = round(inexact(x \circ y))
\]
where $x$ and $y$ are FP numbers, $\circ$ is an arithmetic operation, and $inexact$ is a random perturbation defined at a given virtual precision:
\[
inexact(x)=x+2^{e_{x}-t}\xi
\]
where \(e_{x}\) is the exponent in the FP representation of \(x\), \(t\) is the virtual precision, and \(\xi\) is a random uniform variable of \((-\frac{1}{2},\frac{1}{2})\).
To measure numerical uncertainty, we applied a perturbation of 1 ulp (unit of least precision, a.k.a the spacing between two consecutive FP numbers), which corresponds to a virtual precision of $t=24$ bits for single-precision and $t=53$ bits for double-precision.

We applied RR to the CNN models using Verrou~\cite{fevotte2016verrou}~\cite{verrou-url}, a tool that implements MCA through dynamic binary instrumentation with Valgrind~\cite{nethercote2007valgrind}, without needing to modify or recompile the source code.  We instrumented the entire executables with RR, additionally using Verrou's custom libmath implementation named Interlibmath to avoid incorrect random perturbations in mathematical functions. We applied RR to FreeSurfer using ``fuzzy libmath"~\cite{salari2021accurate}, a version of the GNU mathematical library instrumented with the Verificarlo~\cite{denis2015verificarlo} compiler following the same principle as Verrou's Interlibmath instrumentation.

\subsection{Numerical Uncertainty Metrics}


We quantified numerical uncertainty by calculating the number of significant bits across multiple independent RR samples. The number of significant bits is informally defined as the number of bits in common between RR samples for a given FP value. We estimated the number of significant bits using the general non-parametric method described in~\cite{sohier2021confidence} and implemented in the \texttt{significant\_digits} package~\cite{significantdigits-url}.
Given an RR sample $X_i$ ($i \leq n$), this method computes the significance  $S^k_i$ of the $k^{th}$ bit in the mantissa of $X_i$ as:
\[
S^k_i = \mathbbm{1}_{|Z_i |<2^{-k}}
\]
where $Z_i=X_i-x_\mathrm{IEEE}$ and $x_\mathrm{IEEE}$ is the unperturbed result computed with IEEE.
The $k^{th}$ bit in the mantissa is considered significant if the absolute value of $Z_i$ is less than $2^{-k}$. The number of significant bits across samples, $\hat{s_b}$, is then obtained as the maximal bit index that is significant for all samples:
\[\hat{s_{b}} = \textnormal{max} \left\{ k \in \llbracket 1, m \rrbracket \textnormal{ such that } \forall i \in \llbracket 1, n \rrbracket,\  S^{k}_{i}=\mathbbm{1}\right\} 
\]
where $m$ is the size of the mantissa, i.e., 53 bits for double precision numbers and 24 bits for single precision numbers.
A value of 0 significant bits means that $X$ bears no information while a value of $m$ means that it has maximal information given the FP format used.
The difference between the maximal and achieved values quantifies the information loss resulting from numerical uncertainty.

The number of significant bits is a versatile metric that applies to any program that produces results encoded as FP values. 
This is, however, not the case of segmentation tools that generally produce categorical variables encoded as integers representing segmentation labels 
despite the use of intermediate FP operations.
Therefore, 
in order to assess the impact of stochastic rounding in these intermediate FP operations
we used the minimum Sørensen-Dice scores computed pairwise across RR samples as uncertainty metric for segmentations. In addition, to have a more local uncertainty metric for segmentation results, we defined an entropy metric at each voxel:
\begin{equation}
    \label{eq:entropy}
    E = -\sum_{i=1}^{r}p_i\ln p_i
\end{equation}
where $r$ is the number of segmented regions, and $p_i$ is the probability of region $i$ at the voxel, computed across $n$ Random Rounding samples. 
The entropy is 0 when the voxel is labeled with the same region across all RR samples, and it is maximal when the voxel is labeled differently for each RR sample.

\subsection{Non-Linear Registration}


\paragraph{SynthMorph}\cite{hoffmann2021synthmorph} is a 3D convolutional U-Net~\cite{ronneberger2015u} that performs non-linear image registration robustly across MRI contrasts. The encoding section of the U-Net consists of 4 convolutional blocks, 
while the decoding section consists of 3 blocks 
with a skip connection to its respective encoding block. SynthMorph was trained from synthetic label maps created from geometric shapes from which images were generated with various contrasts, deformations, and artifacts. A contrast-invariant loss that measures label overlap was used to optimize model performance. SynthMorph's registration accuracy was shown to outperform state-of-the-art registration methods both within and across contrasts. We applied the SynthMorph ``sm-brains" pre-trained model available on GitHub~\cite{synthmorph-github} to the linearly registered image produced by FreeSurfer recon-all (output of recon-all's \texttt{-canorm} step), using the MNI305 atlas as reference image. The subject and atlas images were cropped through visual inspection to match the 160 x 192 x 224 dimension required by the model, and intensities were min-max scaled to [0-1]. We applied Verrou directly to the sm-brains model, by instrumenting the entirety of the model and using Verrou's Interlibmath implementation.
SynthMorph takes a couple of minutes to run, but when instrumented with Verrou, the runtime increased to a span of 2-3 days.

\paragraph{FreeSurfer} ``recon-all"~\cite{fischl2002whole} is a widely-used neuroimaging pipeline that implements spatial normalization to a brain template, whole-brain segmentation, and cortical surface extraction. The non-linear registration algorithm (mri\_ca\_register tool) minimizes through gradient descent an error functional that includes an intensity term, a topology constraining one, a metric preservation term,  a smoothness term, and a label term~\cite{fischl2004sequence}.
We first ran recon-all on all the subjects with steps \texttt{--motioncor --talairach --nuintensitycor --normalization --skullstrip --gcareg --canorm} without RR, to obtain the linear registration used as input of SynthMorph. 
Then we ran the FreeSurfer recon-all step \texttt{--careg} that implements non-linear registration using our FreeSurfer version instrumented with fuzzy libmath. This command typically takes 3 hours to run, but when instrumented with fuzzy libmath, the runtime increased to a span of 6 hours.


\subsection{Whole-Brain Segmentation}


\paragraph{FastSurfer}\cite{henschel2020fastsurfer} is a CNN model that performs whole-brain segmentation, cortical surface reconstruction, fast spherical mapping, and cortical thickness analysis. The FastSurfer CNN is inspired from the QuickNAT model~\cite{roy2019quicknat}, which is composed of three 2D fully convolutional neural networks---each associated with a different 2D slice orientation---that each have the same encoder/decoder U-net architecture with skip connections, unpooling layers and dense connections as QuickNAT.
The FastSurfer segmentations were shown to surpass state-of-the-art methods, as well as being generalizable to unseen datasets and having better test-retest reliability. We used the pre-trained model from FastSurfer available on GitHub~\cite{fastsurfer-github} and we applied Verrou directly to this model, in the same way it was applied to SynthMorph. 
FastSurfer typically takes 30 minutes to run, but when instrumented with Verrou, the runtime increased to a span of 16-17 days.

\paragraph{FreeSurfer} recon-all also implements whole-brain segmentation~\cite{fischl2002whole}, through a maximum a-posteriori estimation of the segmentation based on the non-linear registration of the subject image to an atlas.
Due to time constraints, only the subcortical structures and brain tissues were segmented  by FreeSurfer whereas FastSurfer also segmented cortical structures, therefore a mask was applied to FastSurfer's cortical labels to identify them by the super classes ``Left/Right Cerebral Cortex". Only regions common to both were further analysed (see list in Fig.~\ref{fig:fastsurfer_boxplot}).
Similar to FreeSurfer recon-all's non-linear registration, RR was applied to FreeSurfer recon-all's whole brain segmentation through Verificarlo's fuzzy libmath.
The FreeSurfer recon-all commands \texttt{--motioncor} up to \texttt{--calabel}, the command that specifically performs subcortical segmentation were run. 
Typically, the segmentation takes around 4 hours to complete, but with FreeSurfer instrumented with Verificarlo, the runtime increased to 10-12 hours.

\subsection{Dataset and processing}
We used the Consortium for Reliability and Reproducibility (CoRR) dataset~\cite{zuo2014open}, a multi-centric, open resource aimed to evaluate test-retest reliability and reproducibility. We randomly selected 35 T1-weighted MRIs from 35 different subjects, one from each CoRR acquisition site, and accessed them through Datalad~\cite{halchenko2021datalad}~\cite{corr-url-datalad}. The selected images included a range of image dimensions, voxel resolutions and data types (Appendix~\ref{appendix:data}). 
We excluded 2 subjects that failed linear registration with FreeSurfer recon-all and a third subject that failed segmentation with FastSurfer. 
Each pipeline or model was run in Singularity containers over 10 RR samples from which we measured numerical uncertainty. Due to long computing times induced by Verrou instrumentation ($\approx$ 17 days per subject) we were only able to get 4 RR 
samples for FastSurfer, which we complemented with an IEEE (non-RR) sample conceptually identical to an RR sample. 

We processed the data with SynthMorph and FreeSurfer recon-all on the Narval cluster from \'Ecole de Technologie Sup\'erieure (ETS, Montr\'eal), managed by Calcul Qu\'ebec and The Digital Alliance of Canada which include AMD Rome 7502, AMD Rome 7532, and AMD Milan 7413 CPUs with 48 to 64 physical cores, 249~GB to 4000~GB of RAM and Linux kernel 3.10.
We executed FastSurfer on the slashbin cluster at Concordia University
with 8 $\times$ compute nodes each with an Intel Xeon Gold 6130 CPU, 250~GB of RAM, and Linux kernel 4.18.0-240.1.1.el8\_lustre.x86\_64.

We used FreeSurfer v7.3.1, SynthMorph v0.2, FastSurfer v2.1.1, Fuzzy v0.9.1, and Singularity/Apptainer v1.1. Verrou v3.21.0 was used for FastSurfer, while Verrou v3.20.0 with a special fix available on GitHub~\cite{verrou-fix} was used for SynthMorph due to compatibility issues between the model and Verrou's Interlibmath.
The scripts and Dockerfiles for this experiment can be found on GitHub~\cite{scripts}.


\section{Results}

\begin{figure}[t]
  \centering
  \begin{subfigure}[b]{0.48\linewidth}
    \hspace*{-1cm}                                                           
    \includegraphics[width=1.25\linewidth]{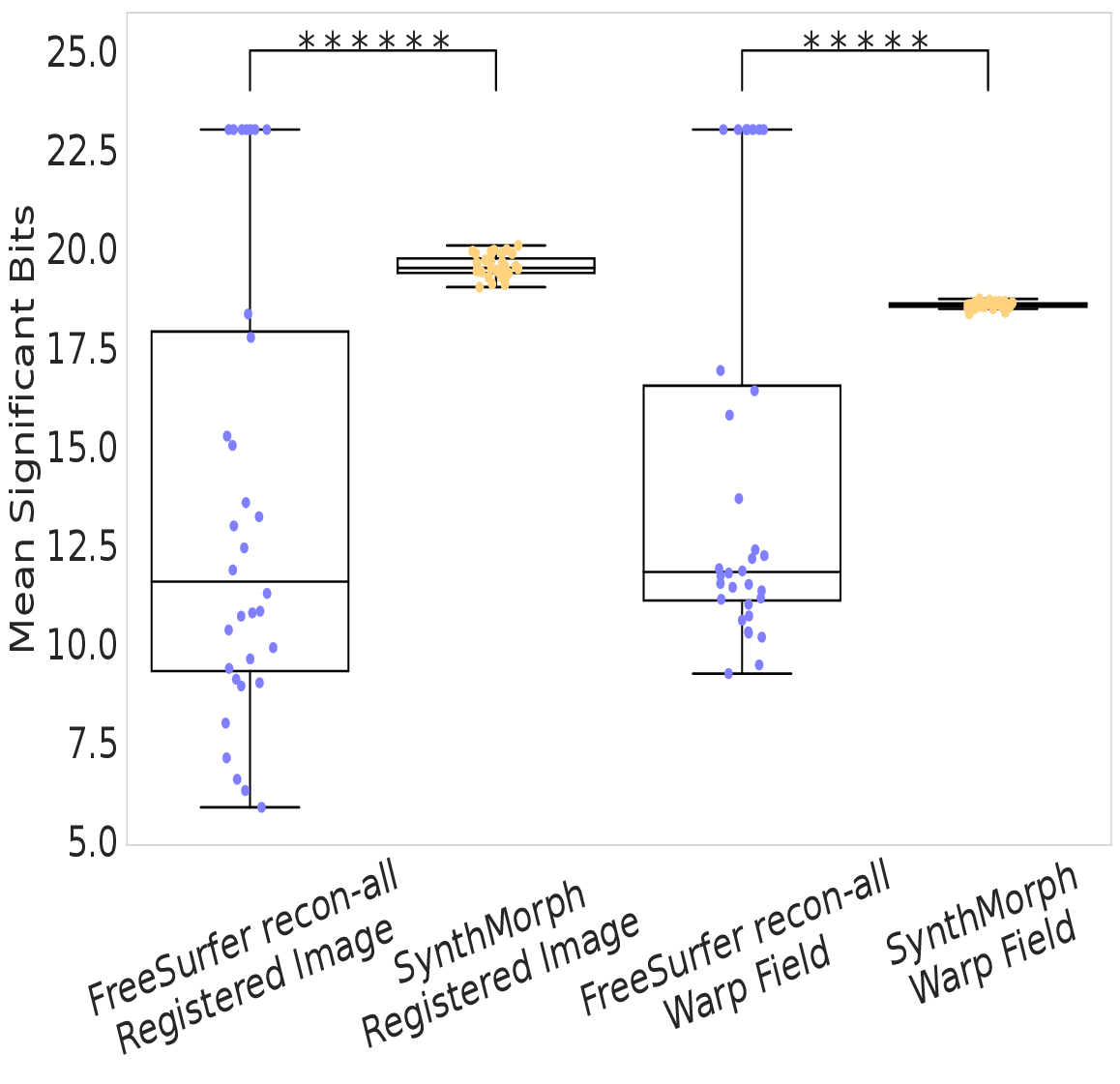}
    \caption{Average numerical uncertainty for n=32 subjects. Each data point represents the voxel-wise mean number of significant bits computed across 10 Random Rounding samples.} 
    \label{sm_boxplot}
  \end{subfigure}
  \hfill
  \hfill
  \begin{subfigure}[b]{0.48\linewidth}
    \centering
  \begin{tabular}{c}
        \includegraphics[width=\linewidth]{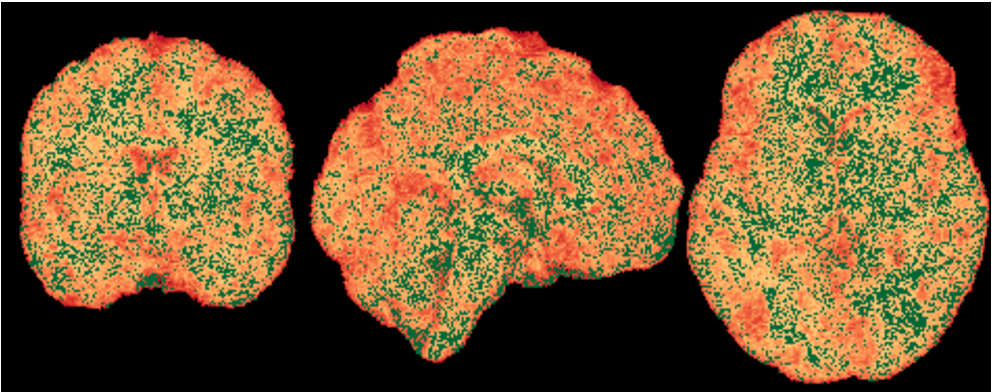} \\
        \scriptsize{FreeSurfer recon-all} \\
        \includegraphics[width=\linewidth]{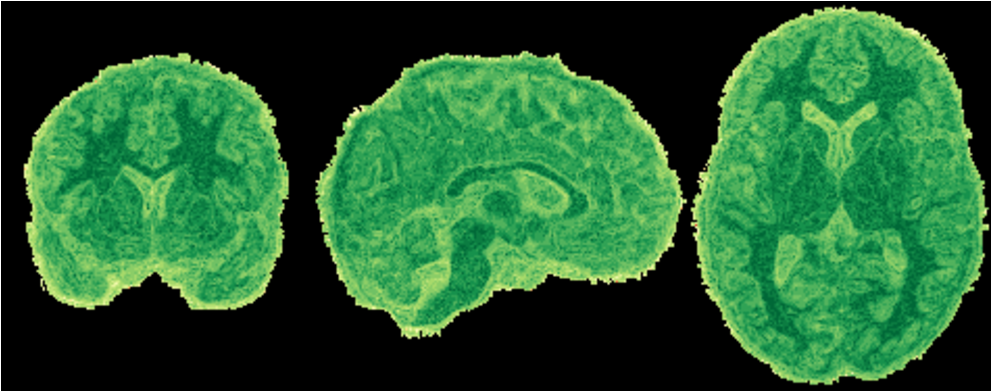} \\
        \scriptsize{SynthMorph} 
  \end{tabular}
  \centering
  \includegraphics[width=\textwidth]{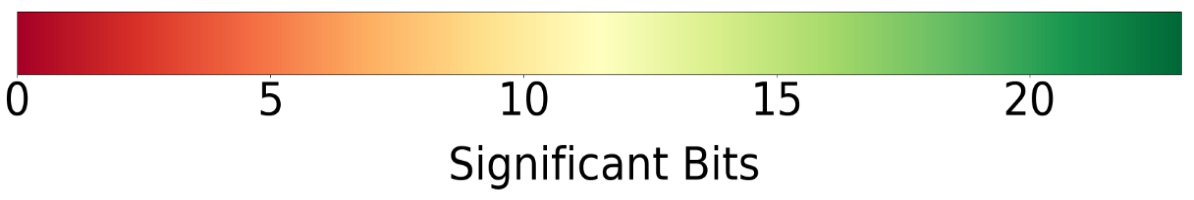}
  \caption{Uncertainty maps for sub-0025555. Uncertainty maps for all the subjects are reported in Appendix~\ref{appendix:non-linear-registration}. Due to required cropping of image dimensions for SynthMorph, the registered images do not have the same dimensions between SynthMorph and FreeSurfer.}
  \label{tbl:table_of_figures}
  \end{subfigure}
  \caption{Numerical uncertainty measured in the non-linearly registered images and warp fields produced by FreeSurfer recon-all and the SynthMorph CNN model. } 
\vspace*{-0.5cm}                                                           

\end{figure}

The numerical uncertainty measured for the SynthMorph CNN model was lower than 
for Freesurfer recon-all (Fig.~\ref{sm_boxplot},
as measured both in the resampled images ($p < 10^{-6}$, two-tailed paired t-test) and in the warp fields ($p < 10^{-5}$) despite only the libmath libraries in FreeSurfer being instrumented in contrast to the entirety of SynthMorph. 
The number of significant bits in warp fields was computed as the average number of significant bits across the x,y and z components.
On average, out of 24 bits available, the SynthMorph resampled image had 19.56 significant bits while FreeSurfer recon-all's had only 13.43 significant bits; the SynthMorph warp field had 18.55 significant bits while FreeSurfer recon-all's had only 14.12 significant bits.
These important differences show a clear superiority of the CNN model compared to FreeSurfer recon-all in terms of numerical uncertainty. 
Moreover, we also observed a larger variability of the numerical uncertainty across subjects in FreeSurfer recon-all compared to SynthMorph. 



The differences in average numerical uncertainty observed between FreeSurfer and SynthMorph were confirmed by visual inspection of the non-linearly registered images and warp fields (Fig.~\ref{tbl:table_of_figures}).
The numerical uncertainty in registered images was structurally consistent, with higher uncertainty in the gray matter and at the border of the brain than in the white matter, both for SynthMorph and for FreeSurfer recon-all. The numerical uncertainty in warp fields exhibited interesting structural patterns that would benefit from further investigation. 

\begin{figure}
  \centering
  \begin{subfigure}[b]{\linewidth}
    \hspace*{-1cm}                                                           
    \includegraphics[width=1.25\textwidth]{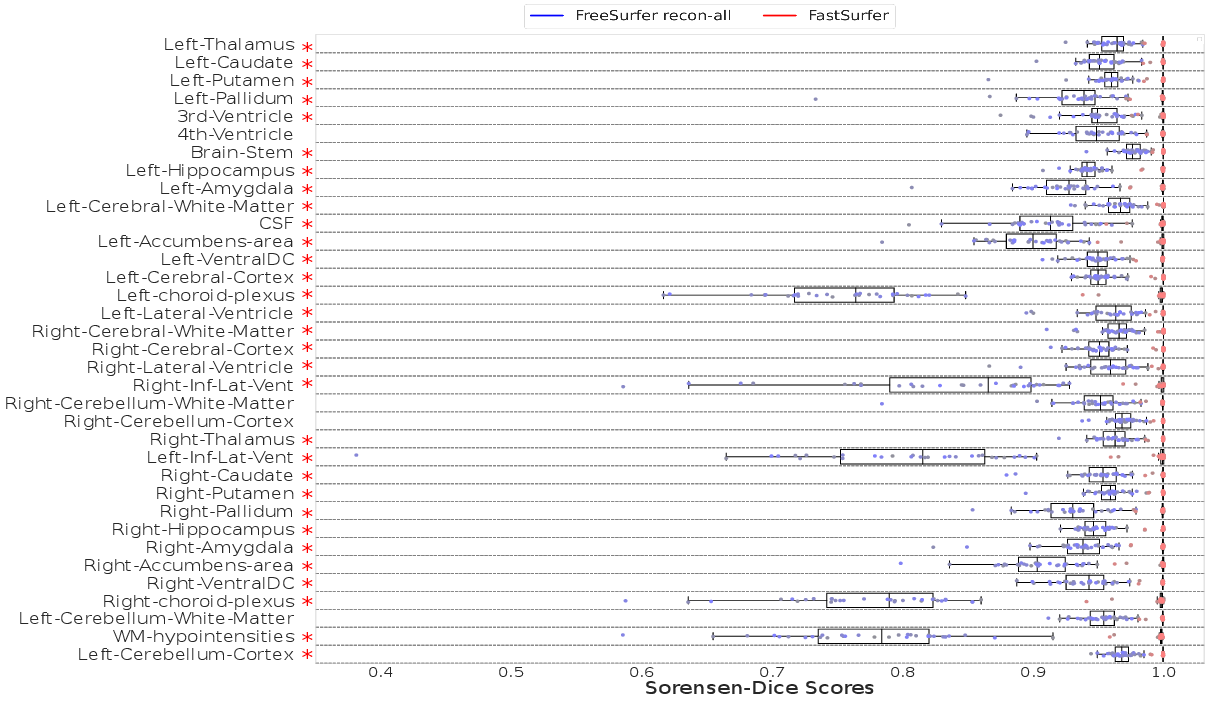}
    \caption{Sørensen-Dice score comparison across FreeSurfer recon-all and FastSurfer for n=33 subjects. Each data point represents the minimum Sørensen-Dice score across all pairs of 5 Random Rounding samples for a given subject. \color{red}*\color{black}\xspace indicates significant differences between FastSurfer and FreeSurfer recon-all segmentations for the region ($p < 0.001$, two-tailed paired t-test with Bonferroni correction).} 
    \label{fig:fastsurfer_boxplot}
  \end{subfigure}
  \begin{subfigure}[b]{\linewidth}
    \centering
    \vspace*{0.5cm}
  \begin{tabular}{c c}
        FreeSurfer recon-all & FastSurfer \\
        \includegraphics[width=0.5\linewidth]{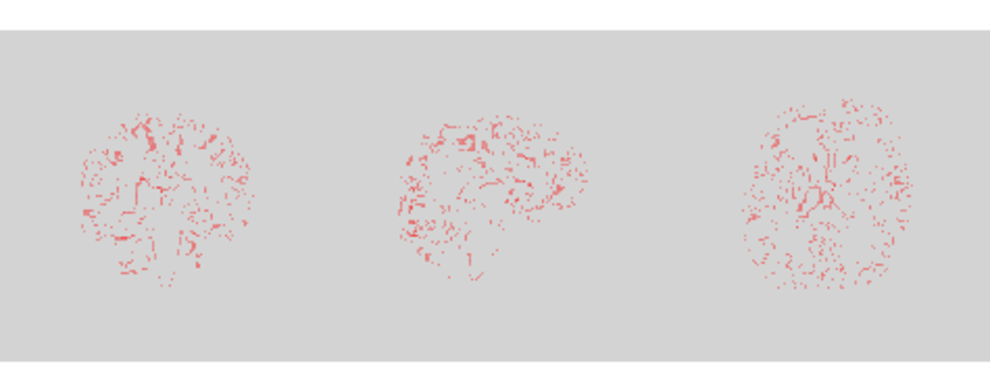} & \includegraphics[width=0.5\linewidth]{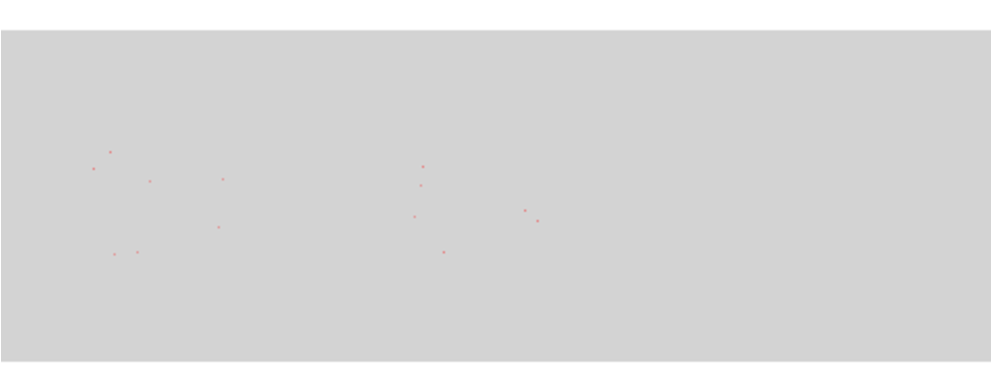} \\
  \end{tabular}
  \centering
  \includegraphics[width=0.6\textwidth]{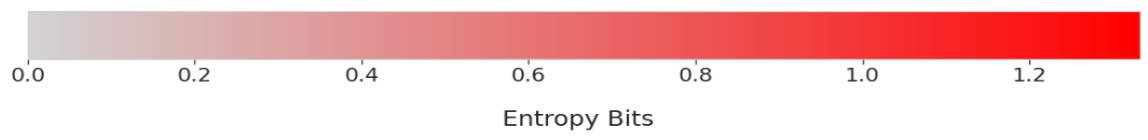}
  \caption{Entropy maps for sub-0027012, computed across r=35 regions and n=5 Random Samples (see Eq.~\ref{eq:entropy}). Entropy maps for all the subjects are reported in Appendix~\ref{appendix:segmentation}.}
  \label{tbl:entropy}
  \end{subfigure}
  \caption{Numerical uncertainty measured in the segmentations produced by FreeSurfer recon-all and the FastSurfer CNN model. } 
\end{figure}

The numerical uncertainty of FastSurfer segmentations was significantly lower than for FreeSurfer recon-all in 31 out of 35 brain regions (Fig.~\ref{fig:fastsurfer_boxplot},
with very substantial differences in some regions.
Here again, a larger variability was observed in FreeSurfer recon-all segmentations than in FastSurfer segmentations.
Overall, FastSurfer averages a Sørensen-Dice score of 0.99 across all regions, while FreeSurfer is at 0.92.
The differences in Sørensen-Dice scores observed between FreeSurfer recon-all and FastSurfer were confirmed in local entropy maps (Fig.~\ref{tbl:entropy} where we visually noted a substantial discrepancy between both methods.
For FreeSurfer recon-all, clusters of non-zero entropy values were observed across the brain, whereas for FastSurfer non-zero entropy values were limited to scattered voxels. 
The entropy maps, in addition to visual inspection, confirm that, despite the relatively high average Sørensen-Dice scores,  FreeSurfer recon-all exhibited variability identifying the edges of subcortical structures, while FastSurfer remained certain in its segmentations.

\newcommand{\addpic}[3]{
  \includegraphics[width=\linewidth]{figures3/#2_#3/#1.eps}
}

\newcommand{\addimg}[3]{
  \includegraphics[width=\linewidth]{figures3/#2_#3/#1_sigmap.eps}
}
\newcolumntype{M}[1]{>{\centering\arraybackslash}m{#1}}


\section{Conclusion}

The numerical uncertainty measured in CNN models SynthMorph and FastSurfer was substantially lower than in FreeSurfer recon-all, amounting to differences in the order of 4 to 6 significant bits in non-linearly registered images, and of up to 0.4 Sørensen-Dice score values in segmentations. 
We believe that the high numerical uncertainty observed in FreeSurfer recon-all compared to CNN models results from the use of numerical optimization techniques in FreeSurfer recon-all while CNN models only involve low-dimensional convolutions, max-pooling operators, and simple activation functions. The low numerical uncertainty found in CNN models is consistent with previous observations in the very different task of protein function prediction~\cite{pepe2022numerical}. The numerical uncertainty found in FreeSurfer recon-all is also consistent with previous observations on FreeSurfer recon-all non-linear registration ~\cite{salari2021accurate} and segmentation~\cite{salari2020file}.

Our results suggest that neuroimaging CNN models are significantly more robust to small numerical perturbations than traditional image processing approaches. Therefore, we expect CNN results to be more reproducible across execution environments than traditional image processing approaches, implying better portability across software and hardware systems. 

Our results report on the numerical uncertainty resulting from CNN \emph{inference}, which is a relevant proxy for the uncertainty experienced by model end-users across different execution environments. However, the numerical uncertainty resulting from CNN \emph{training} was not measured in our experiments. We speculate that some of the numerical uncertainty observed in FreeSurfer recon-all results is intrinsic to the problems of subject-to-template non-linear registration and whole-brain segmentation, and should therefore manifest during CNN training. Mathematically, training CNN models involves numerical optimization in high-dimensional spaces, which we expect to be less numerically stable than CNN inference, and comparably stable to FreeSurfer recon-all. Should this assumption be accurate, the numerical uncertainty of predictions made by a sample of CNN models trained with Random Rounding should be substantial, which we plan to leverage in our future work by building efficient ensemble models capturing the numerical variability associated with non-linear registration or segmentation, possibly resulting in improved predictions. 

\section{Acknowledgements}
Computations were made on the Narval and B\'eluga supercomputers from \'Ecole de Technologie Sup\'erieure (ETS, Montr\'eal), managed by Calcul Qu\'ebec and The Digital Alliance of Canada. The operation of these supercomputers are funded by the Canada Foundation for Innovation (CFI), le Ministère de l’Économie, des Sciences et de l’Innovation du Québec (MESI) and le Fonds de recherche du Québec – Nature et technologies (FRQ-NT).

\bibliographystyle{IEEEtran}
\bibliography{paper}

\clearpage

\appendix 

\section{Image parameters}
\label{appendix:data}

\begin{table}[ht!]

\centering
\begin{tabular}{|c c c c c|} 
 \hline
 Subject & Image Dimension & Voxel Resolution & Data Type & Processing Status \\ [0.5ex] 
 \hline\hline
sub-0027393 & (256, 256, 220) & (0.94, 0.94, 1.00) & float32 & Linear Registration Fail \\
sub-0026196 & (176, 256, 256) & (1.00, 0.98, 0.98) & int16 & Success \\
sub-0026044 & (176, 256, 256) & (1.00, 0.98, 0.98) & float32 & Success \\
sub-0027174 & (176, 256, 256) & (1.00, 0.98, 0.98) & float32 & Success \\
sub-0027268 & (256, 256, 156) & (1.00, 1.00, 1.00) & int16 & Success \\
sub-0027294 & (128, 256, 256) & (1.33, 1.00, 1.00) & float32 & Success \\
sub-0026036 & (160, 480, 512) & (1.20, 0.50, 0.50) & float32 & FastSurfer Segmentation Fail \\
sub-0026017 & (256, 256, 224) & (1.00, 1.00, 1.00) & float32 & Success \\
sub-0025507 & (128, 256, 256) & (1.33, 1.00, 1.00) & float32 & Success \\
sub-0025879 & (144, 256, 256) & (1.33, 1.00, 1.00) & float32 & Success \\
sub-0025350 & (256, 256, 220) & (0.94, 0.94, 1.00) & float32 & Success \\
sub-0025930 & (176, 256, 256) & (1.00, 1.00, 1.00) & float32 & Success \\
sub-0027214 & (176, 256, 256) & (1.00, 0.98, 0.98) & float32 & Success \\
sub-0025555 & (128, 256, 256) & (1.33, 1.00, 1.00) & int16 & Success \\
sub-0025436 & (176, 256, 256) & (1.00, 0.98, 0.98) & float32 & Success \\
sub-0025993 & (176, 256, 256) & (1.00, 1.00, 1.00) & float32 & Success \\
sub-0026175 & (176, 240, 256) & (1.00, 1.00, 1.00) & int16 & Success \\
sub-0025248 & (208, 256, 176) & (1.00, 1.00, 1.00) & float32 & Success \\
sub-0025531 & (160, 240, 256) & (1.20, 0.94, 0.94) & float32 & Success \\
sub-0025011 & (128, 256, 256) & (1.33, 1.00, 1.00) & float32 & Success \\
sub-0027012 & (192, 256, 256) & (1.00, 1.00, 1.00) & float32 & Success \\
sub-0027110 & (176, 256, 256) & (1.00, 1.00, 1.00) & int16 & Success \\
sub-0025362 & (160, 240, 256) & (0.99, 1.00, 1.00) & float32 & Success \\
sub-0025604 & (256, 256, 170) & (1.00, 1.00, 1.00) & float32 & Success \\
sub-0027191 & (176, 256, 256) & (1.00, 0.98, 0.98) & float32 & Success \\
sub-0003002 & (176, 256, 256) & (1.00, 1.00, 1.00) & int16 & Success \\
sub-0026055 & (176, 256, 256) & (1.00, 1.00, 1.00) & float32 & Success \\
sub-0025462 & (188, 256, 256) & (1.00, 1.00, 1.00) & float32 & Success \\
sub-0025631 & (176, 256, 256) & (1.00, 1.00, 1.00) & float32 & Success \\
sub-0025598 & (128, 256, 256) & (1.33, 1.00, 1.00) & float32 & Success \\
sub-2842950 & (192, 256, 256) & (1.00, 1.00, 1.00) & int16 & Success \\
sub-0027094 & (128, 256, 256) & (1.33, 1.00, 1.00) & int16 & Success \\
sub-0025406 & 160, 240, 256) & (1.00, 1.00, 1.00) & float32 & Success \\
sub-0027434 & (240, 320, 320) & (0.70, 0.70, 0.70) & float32 & Linear Registration Fail \\
sub-0027236 & (176, 256, 256) & (0.99, 1.00, 1.00) & float32 & Success \\
 \hline
\end{tabular}
\caption{Subjects sampled in the CoRR dataset. Subjects listed as failures were excluded from analysis}
\label{table:corr_subjects}
\end{table}

\section{Non-linear Registration Uncertainty Maps for 32 Subjects}
\label{appendix:non-linear-registration}

\begin{tabular}{cM{40mm}M{40mm}M{40mm}}
    \toprule
       Subject  & Linearly Registered MRI & FreeSurfer recon-all & SynthMorph \\
    \midrule
    \multirow{5}{*}{0026196} & \multirow{2}{*}{\includegraphics[width=11em]{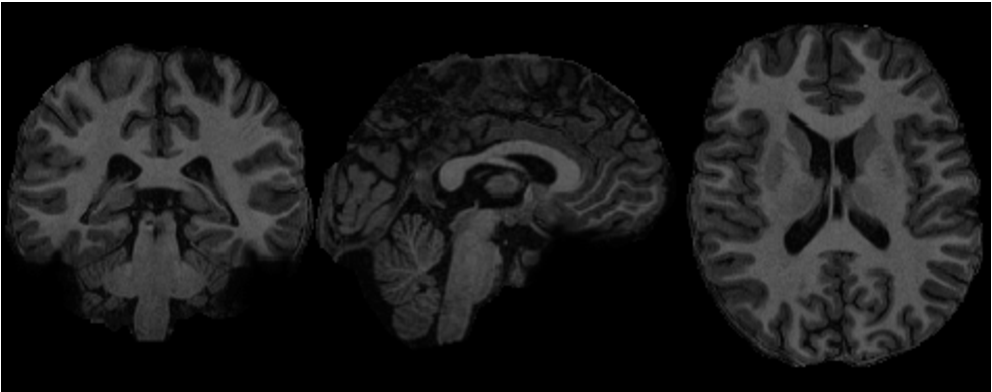}} & \begin{tabular}{cM{40mm}}         \scriptsize{Mean Significant Bits: 5.85} \\ \addimg{sub-0026196}{fs}{reg} \\ \scriptsize{Mean Significant Bits: 10.58} \\ \addimg{sub-0026196}{fs}{warp} \\ \end{tabular} & \begin{tabular}{cM{40mm}} \scriptsize{Mean Significant Bits: 19.01} \\ \addimg{sub-0026196}{sm}{reg} \\ \scriptsize{Mean Significant Bits: 18.48} \\ \addimg{sub-0026196}{sm}{warp} \\ \end{tabular} \\
    
    \multirow{5}{*}{0026044} & \multirow{2}{*}{\includegraphics[width=11em]{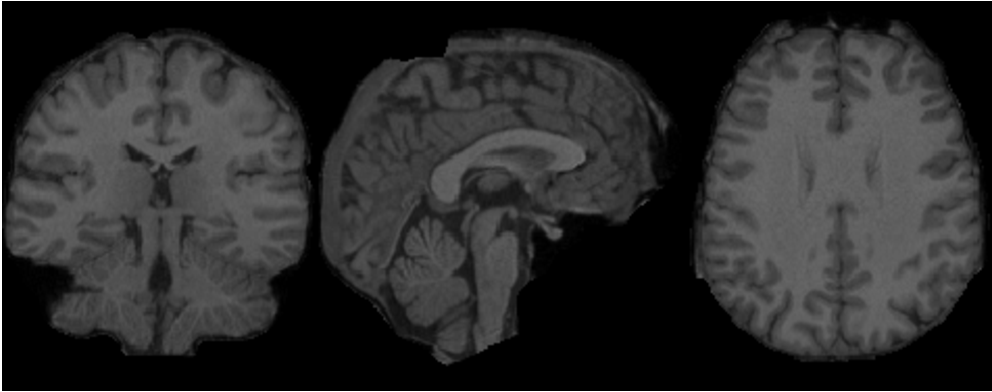}} & \begin{tabular}{cM{40mm}}         \scriptsize{Mean Significant Bits: 6.27} \\ \addimg{sub-0026044}{fs}{reg} \\ \scriptsize{Mean Significant Bits: 9.45} \\ \addimg{sub-0026044}{fs}{warp} \\ \end{tabular} & \begin{tabular}{cM{40mm}} \scriptsize{Mean Significant Bits: 19.96} \\ \addimg{sub-0026044}{sm}{reg} \\ \scriptsize{Mean Significant Bits: 18.60} \\ \addimg{sub-0026044}{sm}{warp} \\ \end{tabular} \\

    \multirow{5}{*}{0027174} & \multirow{2}{*}{\includegraphics[width=11em]{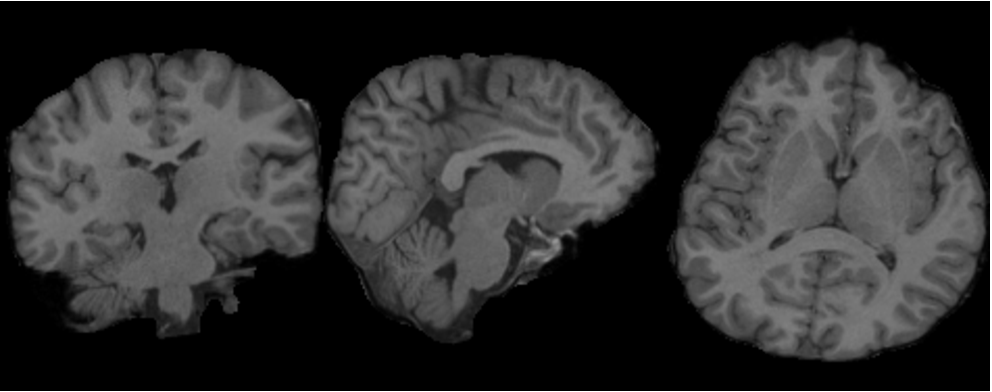}} & \begin{tabular}{cM{40mm}}         \scriptsize{Mean Significant Bits: 6.56} \\ \addimg{sub-0027174}{fs}{reg} \\ \scriptsize{Mean Significant Bits: 9.23} \\ \addimg{sub-0027174}{fs}{warp} \\ \end{tabular} & \begin{tabular}{cM{40mm}} \scriptsize{Mean Significant Bits: 19.92} \\ \addimg{sub-0027174}{sm}{reg} \\ \scriptsize{Mean Significant Bits: 18.65} \\ \addimg{sub-0027174}{sm}{warp} \\ \end{tabular} \\

    \multirow{5}{*}{0027268} & \multirow{2}{*}{\includegraphics[width=11em]{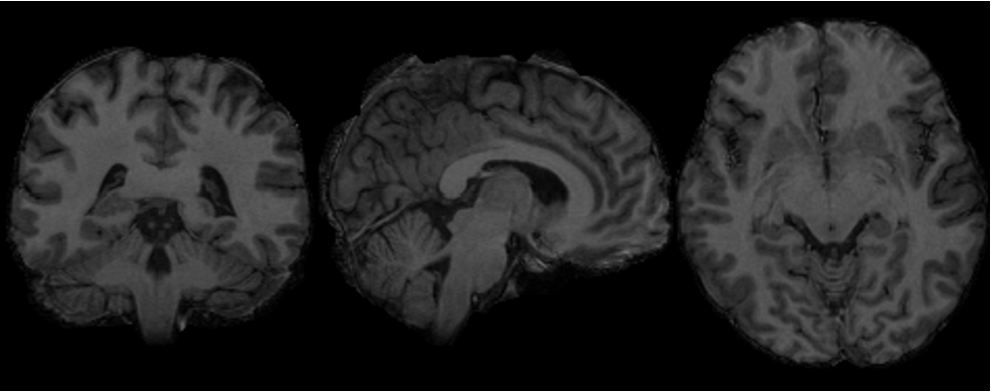}} & \begin{tabular}{cM{40mm}}         \scriptsize{Mean Significant Bits: 7.10} \\ \addimg{sub-0027268}{fs}{reg} \\ \scriptsize{Mean Significant Bits: 11.49} \\ \addimg{sub-0027268}{fs}{warp} \\ \end{tabular} & \begin{tabular}{cM{40mm}} \scriptsize{Mean Significant Bits: 19.48} \\ \addimg{sub-0027268}{sm}{reg} \\ \scriptsize{Mean Significant Bits: 18.57} \\ \addimg{sub-0027268}{sm}{warp} \\ \end{tabular} \\

  \end{tabular}

  \begin{tabular}{cM{40mm}M{40mm}M{40mm}}
    \multirow{5}{*}{0027294} & \multirow{2}{*}{\includegraphics[width=11em]{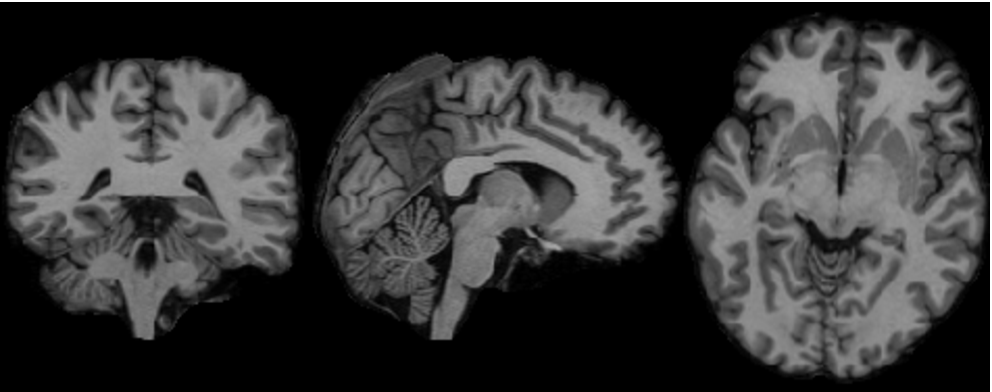}} & \begin{tabular}{cM{40mm}}         \scriptsize{Mean Significant Bits: 7.98} \\ \addimg{sub-0027294}{fs}{reg} \\ \scriptsize{Mean Significant Bits: 10.26} \\ \addimg{sub-0027294}{fs}{warp} \\ \end{tabular} & \begin{tabular}{cM{40mm}} \scriptsize{Mean Significant Bits: 19.54} \\ \addimg{sub-0027294}{sm}{reg} \\ \scriptsize{Mean Significant Bits: 18.71} \\ \addimg{sub-0027294}{sm}{warp} \\ \end{tabular} \\
    
    \multirow{5}{*}{0026017} & \multirow{2}{*}{\includegraphics[width=11em]{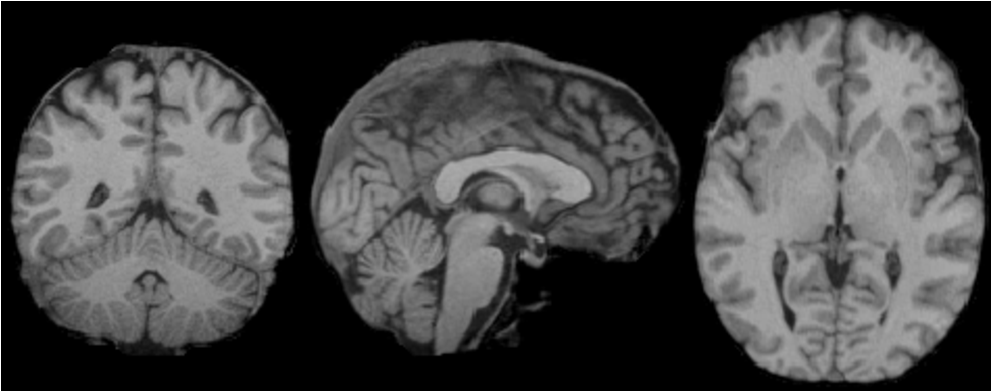}} & \begin{tabular}{cM{40mm}}         \scriptsize{Mean Significant Bits: 8.92} \\ \addimg{sub-0026017}{fs}{reg} \\ \scriptsize{Mean Significant Bits: 11.14} \\ \addimg{sub-0026017}{fs}{warp} \\ \end{tabular} & \begin{tabular}{cM{40mm}} \scriptsize{Mean Significant Bits: 19.87} \\ \addimg{sub-0026017}{sm}{reg} \\ \scriptsize{Mean Significant Bits: 18.49} \\ \addimg{sub-0026017}{sm}{warp} \\ \end{tabular} \\
    
    \multirow{5}{*}{0025507} & \multirow{2}{*}{\includegraphics[width=11em]{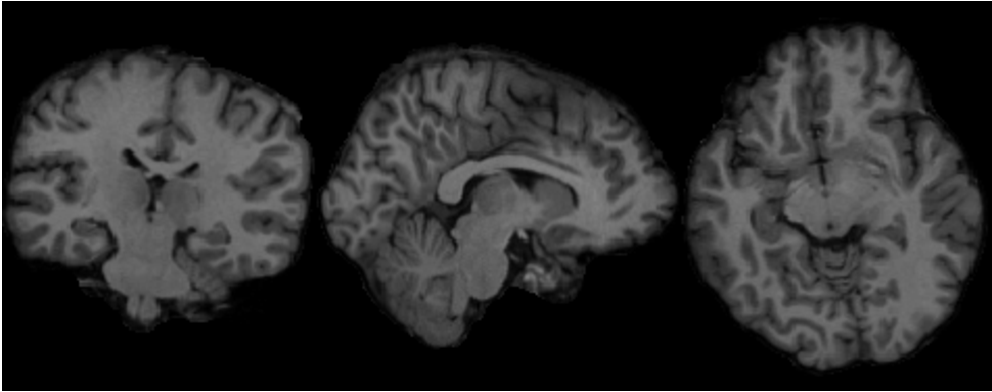}} & \begin{tabular}{cM{40mm}}         \scriptsize{Mean Significant Bits: 9.00} \\ \addimg{sub-0025507}{fs}{reg} \\ \scriptsize{Mean Significant Bits: 11.51} \\ \addimg{sub-0025507}{fs}{warp} \\ \end{tabular} & \begin{tabular}{cM{40mm}} \scriptsize{Mean Significant Bits: 19.65} \\ \addimg{sub-0025507}{sm}{reg} \\ \scriptsize{Mean Significant Bits: 18.65} \\ \addimg{sub-0025507}{sm}{warp} \\ \end{tabular} \\

    \multirow{5}{*}{0025879} & \multirow{2}{*}{\includegraphics[width=11em]{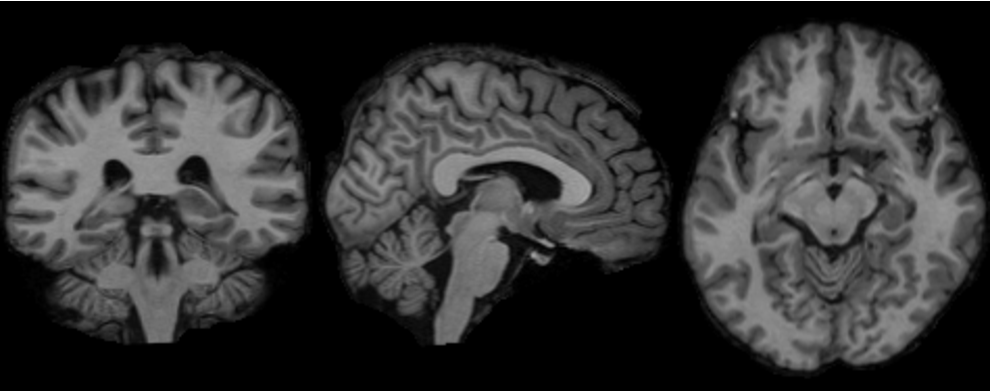}} & \begin{tabular}{cM{40mm}}         \scriptsize{Mean Significant Bits: 9.09} \\ \addimg{sub-0025879}{fs}{reg} \\ \scriptsize{Mean Significant Bits: 10.69} \\ \addimg{sub-0025879}{fs}{warp} \\ \end{tabular} & \begin{tabular}{cM{40mm}} \scriptsize{Mean Significant Bits: 19.53} \\ \addimg{sub-0025879}{sm}{reg} \\ \scriptsize{Mean Significant Bits: 18.57} \\ \addimg{sub-0025879}{sm}{warp} \\ \end{tabular} \\

    \multirow{5}{*}{0025350} & \multirow{2}{*}{\includegraphics[width=11em]{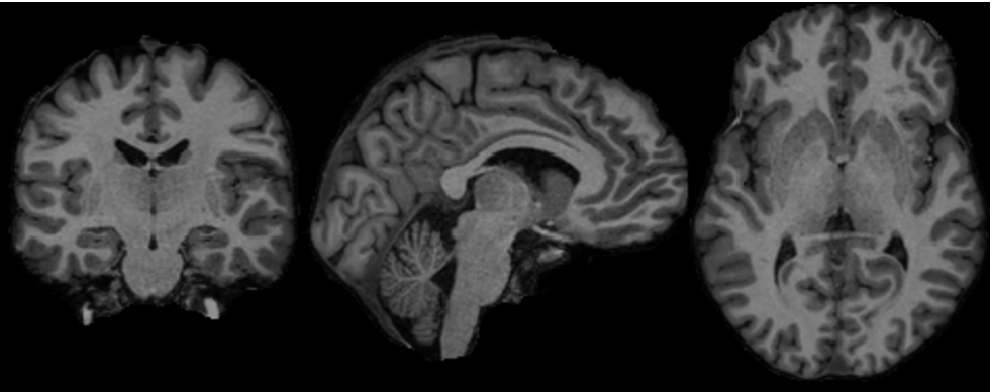}} & \begin{tabular}{cM{40mm}}         \scriptsize{Mean Significant Bits: 9.36} \\ \addimg{sub-0025350}{fs}{reg} \\ \scriptsize{Mean Significant Bits: 11.33} \\ \addimg{sub-0025350}{fs}{warp} \\ \end{tabular} & \begin{tabular}{cM{40mm}} \scriptsize{Mean Significant Bits: 19.38} \\ \addimg{sub-0025350}{sm}{reg} \\ \scriptsize{Mean Significant Bits: 18.34} \\ \addimg{sub-0025350}{sm}{warp} \\ \end{tabular} \\

  \end{tabular}

  \begin{tabular}{cM{40mm}M{40mm}M{40mm}}
    \multirow{5}{*}{0025930} & \multirow{2}{*}{\includegraphics[width=11em]{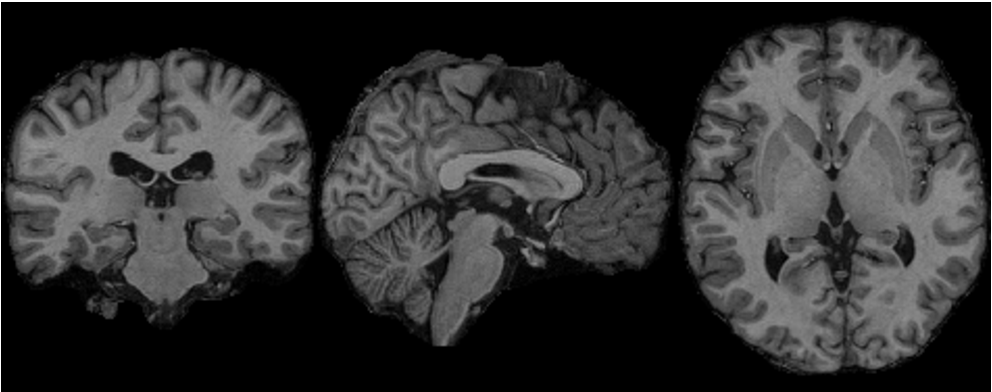}} & \begin{tabular}{cM{40mm}}         \scriptsize{Mean Significant Bits: 9.60} \\ \addimg{sub-0025930}{fs}{reg} \\ \scriptsize{Mean Significant Bits: 11.83} \\ \addimg{sub-0025930}{fs}{warp} \\ \end{tabular} & \begin{tabular}{cM{40mm}} \scriptsize{Mean Significant Bits: 19.37} \\ \addimg{sub-0025930}{sm}{reg} \\ \scriptsize{Mean Significant Bits: 18.51} \\ \addimg{sub-0025930}{sm}{warp} \\ \end{tabular} \\

    \multirow{5}{*}{0027214} & \multirow{2}{*}{\includegraphics[width=11em]{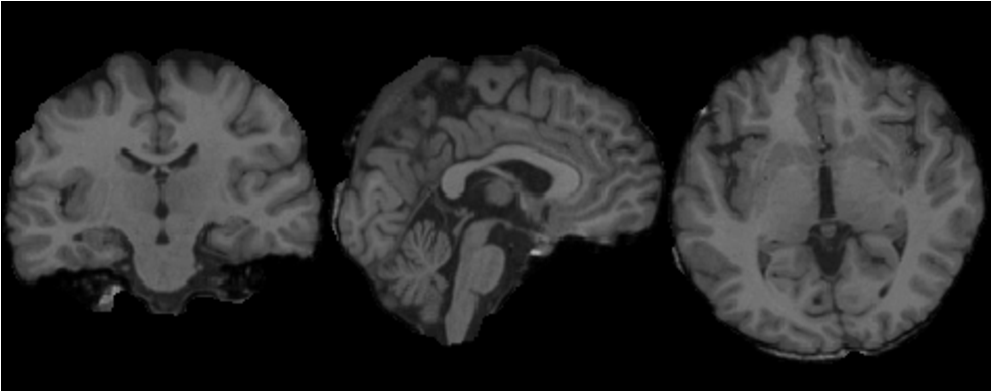}} & \begin{tabular}{cM{40mm}}         \scriptsize{Mean Significant Bits: 9.89} \\ \addimg{sub-0027214}{fs}{reg} \\ \scriptsize{Mean Significant Bits: 10.29} \\ \addimg{sub-0027214}{fs}{warp} \\ \end{tabular} & \begin{tabular}{cM{40mm}} \scriptsize{Mean Significant Bits: 19.86} \\ \addimg{sub-0027214}{sm}{reg} \\ \scriptsize{Mean Significant Bits: 18.53} \\ \addimg{sub-0027214}{sm}{warp} \\ \end{tabular} \\
    
    \multirow{5}{*}{0025555} & \multirow{2}{*}{\includegraphics[width=11em]{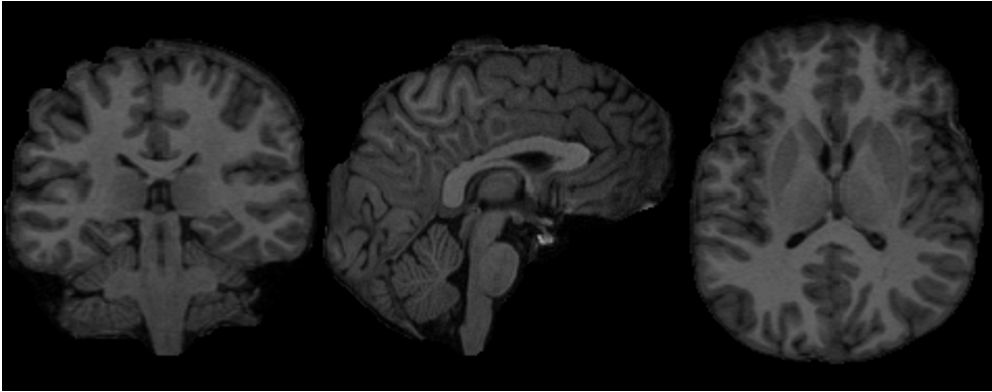}} & \begin{tabular}{cM{40mm}}         \scriptsize{Mean Significant Bits: 10.34} \\ \addimg{sub-0025555}{fs}{reg} \\ \scriptsize{Mean Significant Bits: 11.78} \\ \addimg{sub-0025555}{fs}{warp} \\ \end{tabular} & \begin{tabular}{cM{40mm}} \scriptsize{Mean Significant Bits: 19.53} \\ \addimg{sub-0025555}{sm}{reg} \\ \scriptsize{Mean Significant Bits: 18.47} \\ \addimg{sub-0025555}{sm}{warp} \\ \end{tabular} \\

    \multirow{5}{*}{0025436} & \multirow{2}{*}{\includegraphics[width=11em]{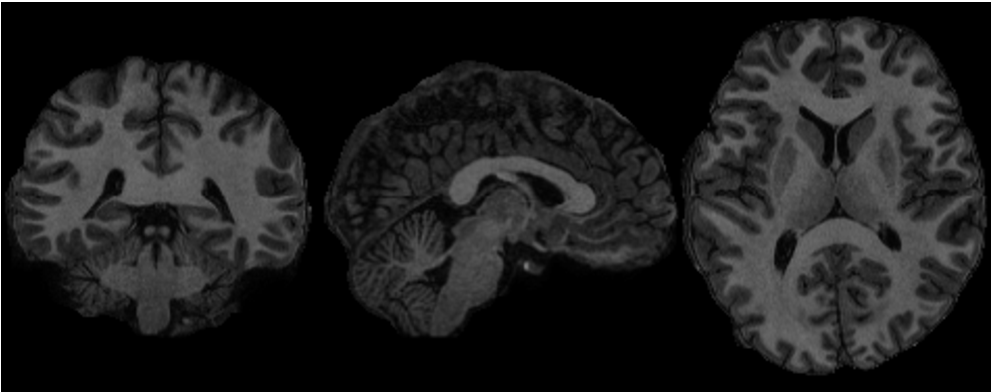}} & \begin{tabular}{cM{40mm}}         \scriptsize{Mean Significant Bits: 10.68} \\ \addimg{sub-0025436}{fs}{reg} \\ \scriptsize{Mean Significant Bits: 13.66} \\ \addimg{sub-0025436}{fs}{warp} \\ \end{tabular} & \begin{tabular}{cM{40mm}} \scriptsize{Mean Significant Bits: 19.09} \\ \addimg{sub-0025436}{sm}{reg} \\ \scriptsize{Mean Significant Bits: 18.46} \\ \addimg{sub-0025436}{sm}{warp} \\ \end{tabular} \\

    \multirow{5}{*}{0025993} & \multirow{2}{*}{\includegraphics[width=11em]{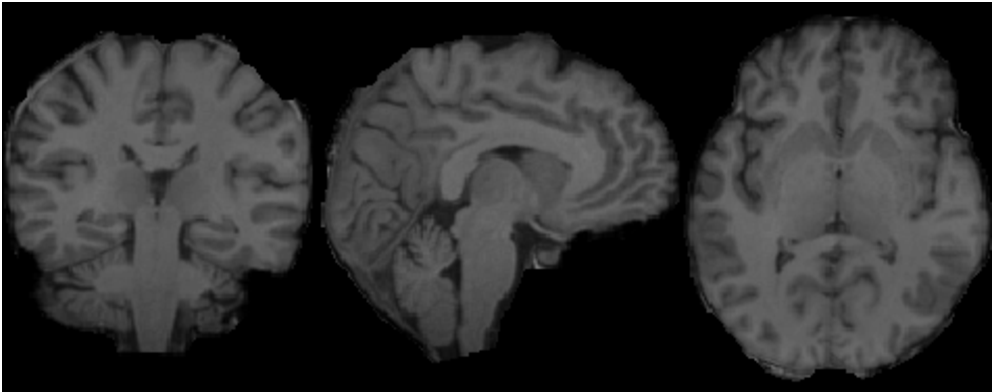}} & \begin{tabular}{cM{40mm}}         \scriptsize{Mean Significant Bits: 10.77} \\ \addimg{sub-0025993}{fs}{reg} \\ \scriptsize{Mean Significant Bits: 10.15} \\ \addimg{sub-0025993}{fs}{warp} \\ \end{tabular} & \begin{tabular}{cM{40mm}} \scriptsize{Mean Significant Bits: 20.07} \\ \addimg{sub-0025993}{sm}{reg} \\ \scriptsize{Mean Significant Bits: 18.62} \\ \addimg{sub-0025993}{sm}{warp} \\ \end{tabular} \\

  \end{tabular}

    \begin{tabular}{cM{40mm}M{40mm}M{40mm}}
    \multirow{5}{*}{0026175} & \multirow{2}{*}{\includegraphics[width=11em]{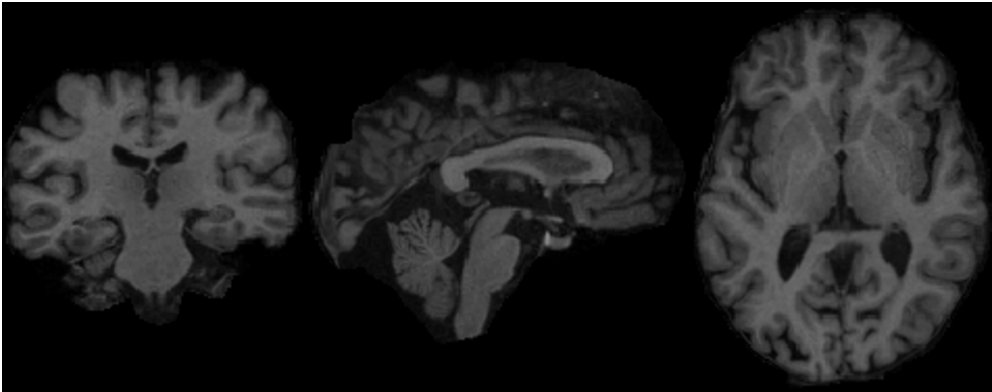}} & \begin{tabular}{cM{40mm}}         \scriptsize{Mean Significant Bits: 10.81} \\ \addimg{sub-0026175}{fs}{reg} \\ \scriptsize{Mean Significant Bits: 11.42} \\ \addimg{sub-0026175}{fs}{warp} \\ \end{tabular} & \begin{tabular}{cM{40mm}} \scriptsize{Mean Significant Bits: 19.37} \\ \addimg{sub-0026175}{sm}{reg} \\ \scriptsize{Mean Significant Bits: 18.54} \\ \addimg{sub-0026175}{sm}{warp} \\ \end{tabular} \\

    \multirow{5}{*}{0025248} & \multirow{2}{*}{\includegraphics[width=11em]{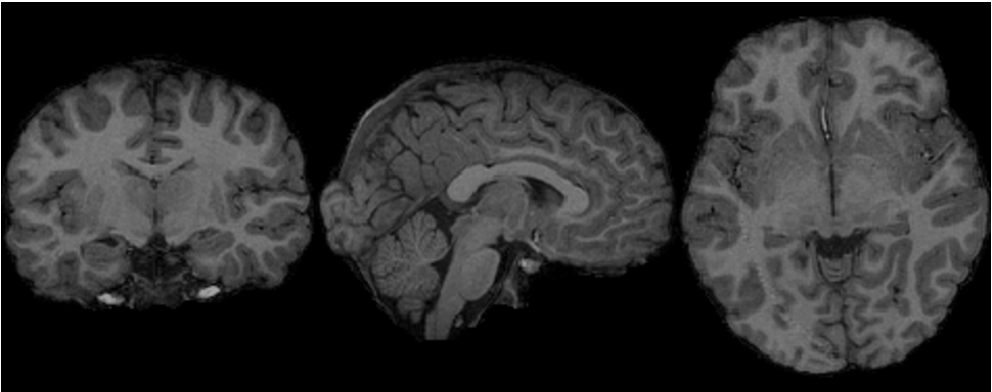}} & \begin{tabular}{cM{40mm}}         \scriptsize{Mean Significant Bits: 11.26} \\ \addimg{sub-0025248}{fs}{reg} \\ \scriptsize{Mean Significant Bits: 12.14} \\ \addimg{sub-0025248}{fs}{warp} \\ \end{tabular} & \begin{tabular}{cM{40mm}} \scriptsize{Mean Significant Bits: 19.51} \\ \addimg{sub-0025248}{sm}{reg} \\ \scriptsize{Mean Significant Bits: 18.56} \\ \addimg{sub-0025248}{sm}{warp} \\ \end{tabular} \\
    
    \multirow{5}{*}{0025531} & \multirow{2}{*}{\includegraphics[width=11em]{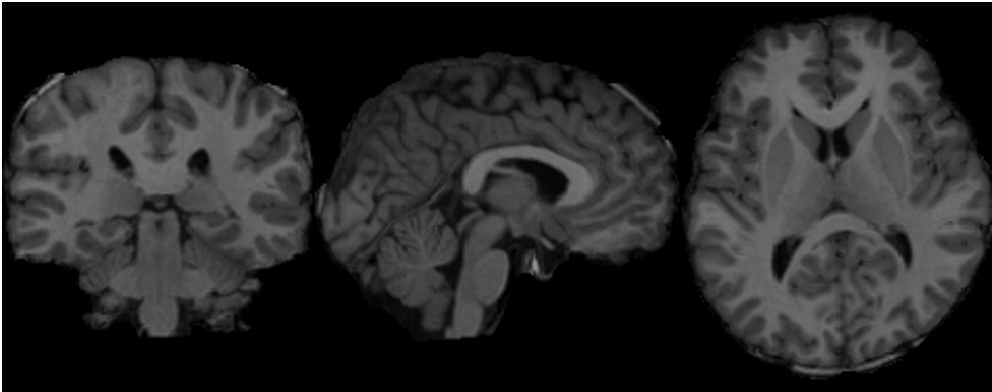}} & \begin{tabular}{cM{40mm}}         \scriptsize{Mean Significant Bits: 11.85} \\ \addimg{sub-0025531}{fs}{reg} \\ \scriptsize{Mean Significant Bits: 10.99} \\ \addimg{sub-0025531}{fs}{warp} \\ \end{tabular} & \begin{tabular}{cM{40mm}} \scriptsize{Mean Significant Bits: 19.68} \\ \addimg{sub-0025531}{sm}{reg} \\ \scriptsize{Mean Significant Bits: 18.59} \\ \addimg{sub-0025531}{sm}{warp} \\ \end{tabular} \\

    \multirow{5}{*}{0025011} & \multirow{2}{*}{\includegraphics[width=11em]{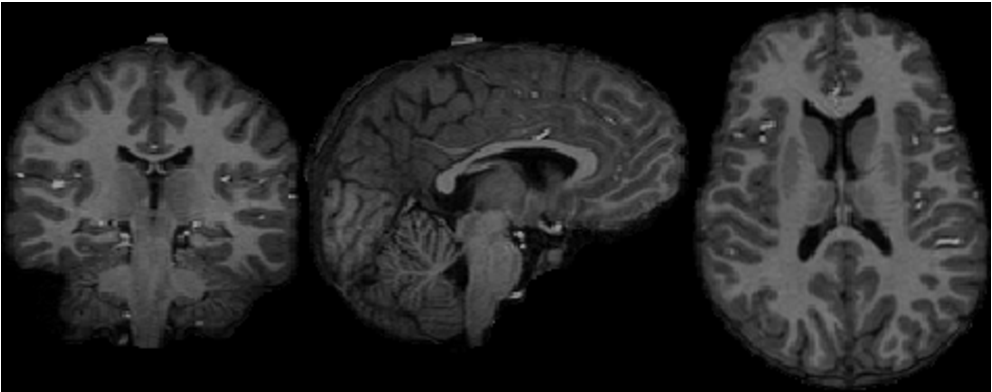}} & \begin{tabular}{cM{40mm}}         \scriptsize{Mean Significant Bits: 12.41} \\ \addimg{sub-0025011}{fs}{reg} \\ \scriptsize{Mean Significant Bits: 12.37} \\ \addimg{sub-0025011}{fs}{warp} \\ \end{tabular} & \begin{tabular}{cM{40mm}} \scriptsize{Mean Significant Bits: 19.27} \\ \addimg{sub-0025011}{sm}{reg} \\ \scriptsize{Mean Significant Bits: 18.69} \\ \addimg{sub-0025011}{sm}{warp} \\ \end{tabular} \\

    \multirow{5}{*}{0027012} & \multirow{2}{*}{\includegraphics[width=11em]{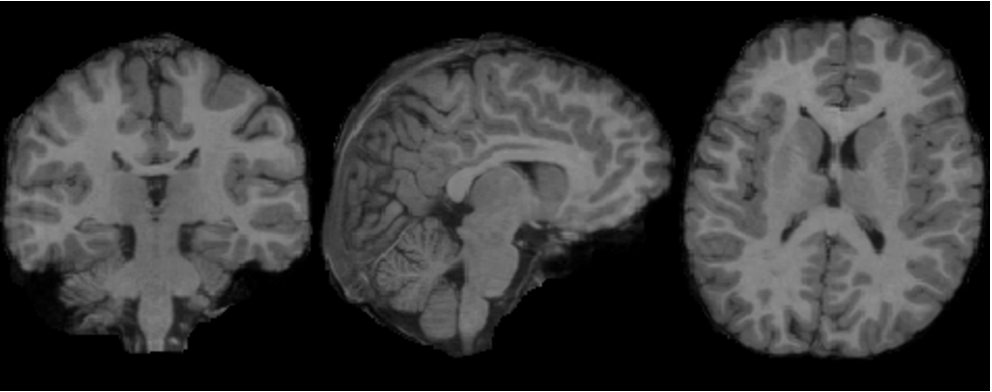}} & \begin{tabular}{cM{40mm}}         \scriptsize{Mean Significant Bits: 12.97} \\ \addimg{sub-0027012}{fs}{reg} \\ \scriptsize{Mean Significant Bits: 11.88} \\ \addimg{sub-0027012}{fs}{warp} \\ \end{tabular} & \begin{tabular}{cM{40mm}} \scriptsize{Mean Significant Bits: 19.86} \\ \addimg{sub-0027012}{sm}{reg} \\ \scriptsize{Mean Significant Bits: 18.59} \\ \addimg{sub-0027012}{sm}{warp} \\ \end{tabular} \\

  \end{tabular}

      \begin{tabular}{cM{40mm}M{40mm}M{40mm}}
    \multirow{5}{*}{0027110} & \multirow{2}{*}{\includegraphics[width=11em]{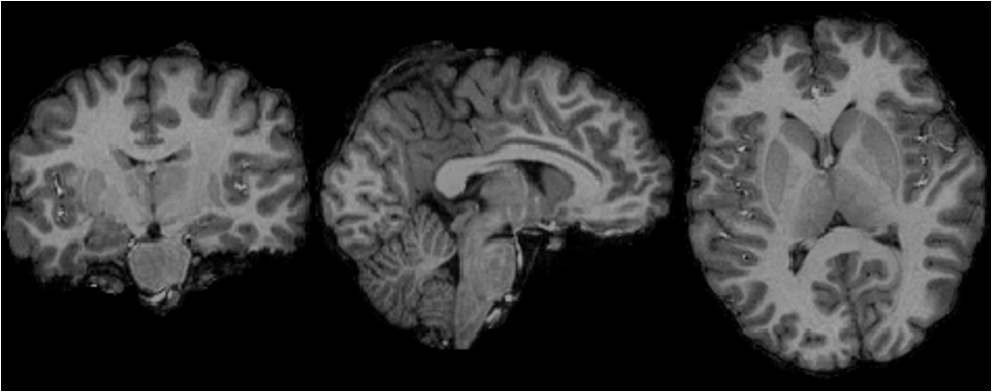}} & \begin{tabular}{cM{40mm}}         \scriptsize{Mean Significant Bits: 13.20} \\ \addimg{sub-0027110}{fs}{reg} \\ \scriptsize{Mean Significant Bits: 16.39} \\ \addimg{sub-0027110}{fs}{warp} \\ \end{tabular} & \begin{tabular}{cM{40mm}} \scriptsize{Mean Significant Bits: 19.27} \\ \addimg{sub-0027110}{sm}{reg} \\ \scriptsize{Mean Significant Bits: 18.52} \\ \addimg{sub-0027110}{sm}{warp} \\ \end{tabular} \\

    \multirow{5}{*}{0025362} & \multirow{2}{*}{\includegraphics[width=11em]{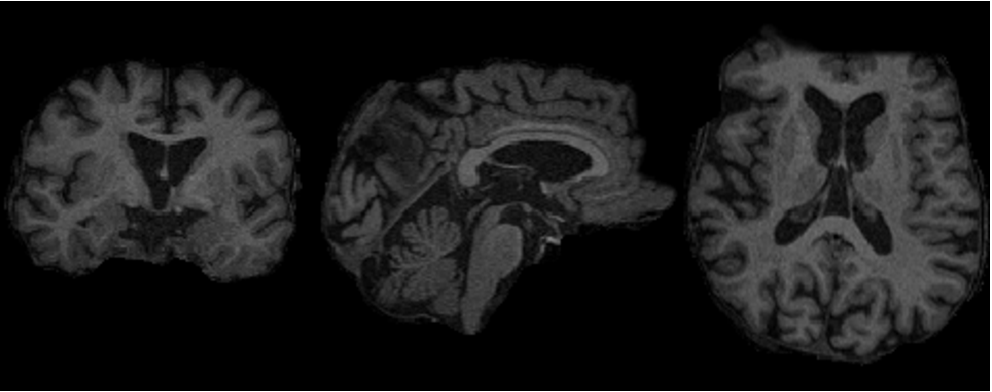}} & \begin{tabular}{cM{40mm}}         \scriptsize{Mean Significant Bits: 13.56} \\ \addimg{sub-0025362}{fs}{reg} \\ \scriptsize{Mean Significant Bits: 11.71} \\ \addimg{sub-0025362}{fs}{warp} \\ \end{tabular} & \begin{tabular}{cM{40mm}} \scriptsize{Mean Significant Bits: 19.07} \\ \addimg{sub-0025362}{sm}{reg} \\ \scriptsize{Mean Significant Bits: 18.61} \\ \addimg{sub-0025362}{sm}{warp} \\ \end{tabular} \\
    
    \multirow{5}{*}{0025604} & \multirow{2}{*}{\includegraphics[width=11em]{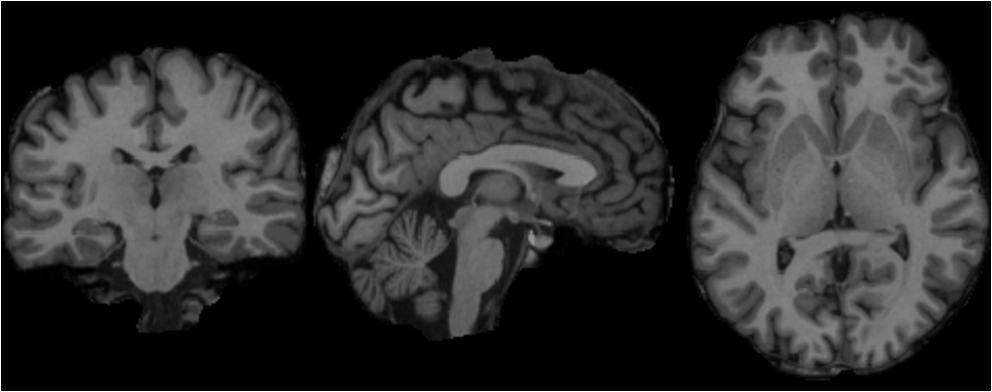}} & \begin{tabular}{cM{40mm}}         \scriptsize{Mean Significant Bits: 15.01} \\ \addimg{sub-0025604}{fs}{reg} \\ \scriptsize{Mean Significant Bits: 11.11} \\ \addimg{sub-0025604}{fs}{warp} \\ \end{tabular} & \begin{tabular}{cM{40mm}} \scriptsize{Mean Significant Bits: 19.63} \\ \addimg{sub-0025604}{sm}{reg} \\ \scriptsize{Mean Significant Bits: 18.52} \\ \addimg{sub-0025604}{sm}{warp} \\ \end{tabular} \\

    \multirow{5}{*}{0027191} & \multirow{2}{*}{\includegraphics[width=11em]{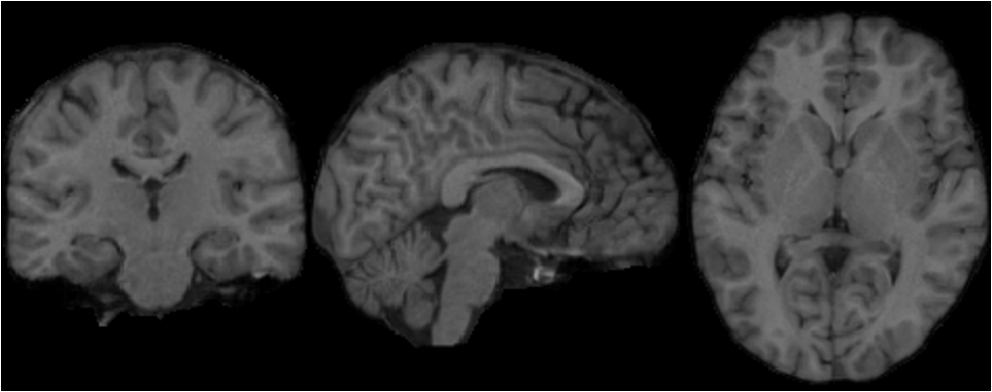}} & \begin{tabular}{cM{40mm}}         \scriptsize{Mean Significant Bits: 15.24} \\ \addimg{sub-0027191}{fs}{reg} \\ \scriptsize{Mean Significant Bits: 12.22} \\ \addimg{sub-0027191}{fs}{warp} \\ \end{tabular} & \begin{tabular}{cM{40mm}} \scriptsize{Mean Significant Bits:19.95} \\ \addimg{sub-0027191}{sm}{reg} \\ \scriptsize{Mean Significant Bits: 18.59} \\ \addimg{sub-0027191}{sm}{warp} \\ \end{tabular} \\

    \multirow{5}{*}{0003002} & \multirow{2}{*}{\includegraphics[width=11em]{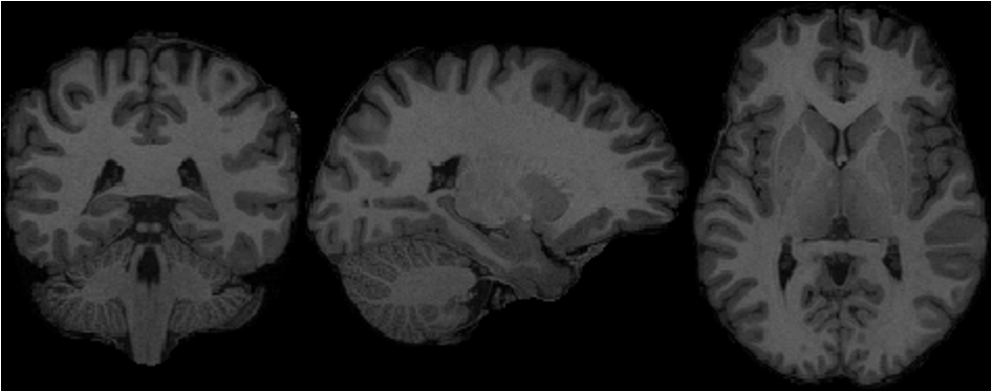}} & \begin{tabular}{cM{40mm}}         \scriptsize{Mean Significant Bits: 17.74} \\ \addimg{sub-0003002}{fs}{reg} \\ \scriptsize{Mean Significant Bits: 16.90} \\ \addimg{sub-0003002}{fs}{warp} \\ \end{tabular} & \begin{tabular}{cM{40mm}} \scriptsize{Mean Significant Bits: 19.40} \\ \addimg{sub-0003002}{sm}{reg} \\ \scriptsize{Mean Significant Bits: 18.60} \\ \addimg{sub-0003002}{sm}{warp} \\ \end{tabular} \\

  \end{tabular}

        \begin{tabular}{cM{40mm}M{40mm}M{40mm}}
    \multirow{5}{*}{0026055} & \multirow{2}{*}{\includegraphics[width=11em]{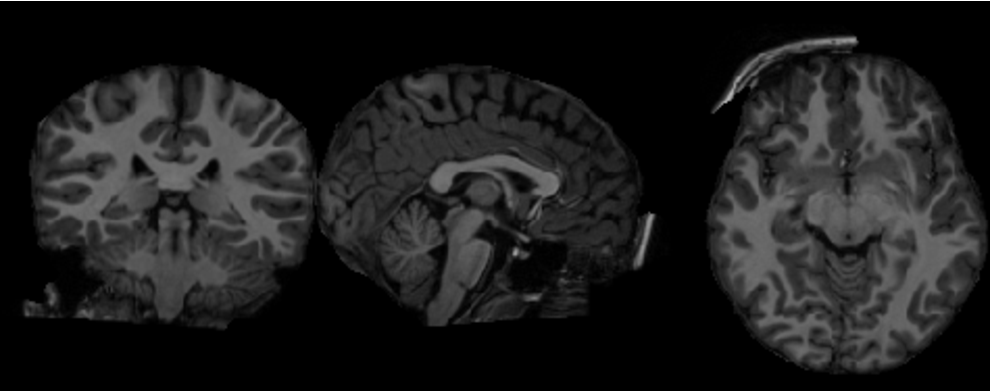}} & \begin{tabular}{cM{40mm}}         \scriptsize{Mean Significant Bits: 18.33} \\ \addimg{sub-0026055}{fs}{reg} \\ \scriptsize{Mean Significant Bits: 15.77} \\ \addimg{sub-0026055}{fs}{warp} \\ \end{tabular} & \begin{tabular}{cM{40mm}} \scriptsize{Mean Significant Bits: 19.23} \\ \addimg{sub-0026055}{sm}{reg} \\ \scriptsize{Mean Significant Bits: 18.66} \\ \addimg{sub-0026055}{sm}{warp} \\ \end{tabular} \\

    \multirow{5}{*}{0025462} & \multirow{2}{*}{\includegraphics[width=11em]{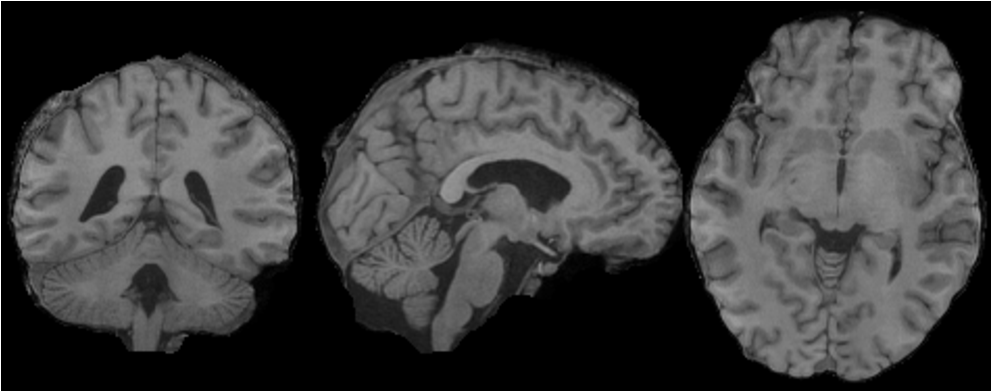}} & \begin{tabular}{cM{40mm}}         \scriptsize{Mean Significant Bits: 23.00} \\ \addimg{sub-0025462}{fs}{reg} \\ \scriptsize{Mean Significant Bits: 23.00} \\ \addimg{sub-0025462}{fs}{warp} \\ \end{tabular} & \begin{tabular}{cM{40mm}} \scriptsize{Mean Significant Bits: 19.91} \\ \addimg{sub-0025462}{sm}{reg} \\ \scriptsize{Mean Significant Bits: 18.37} \\ \addimg{sub-0025462}{sm}{warp} \\ \end{tabular} \\
    
    \multirow{5}{*}{0025631} & \multirow{2}{*}{\includegraphics[width=11em]{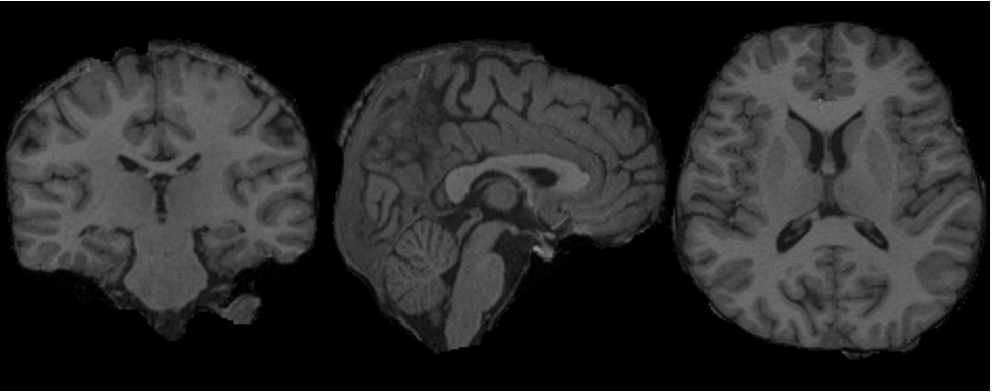}} & \begin{tabular}{cM{40mm}}         \scriptsize{Mean Significant Bits: 23.00} \\ \addimg{sub-0025631}{fs}{reg} \\ \scriptsize{Mean Significant Bits: 23.00} \\ \addimg{sub-0025631}{fs}{warp} \\ \end{tabular} & \begin{tabular}{cM{40mm}} \scriptsize{Mean Significant Bits: 19.70} \\ \addimg{sub-0025631}{sm}{reg} \\ \scriptsize{Mean Significant Bits: 18.58} \\ \addimg{sub-0025631}{sm}{warp} \\ \end{tabular} \\

    \multirow{5}{*}{0025598} & \multirow{2}{*}{\includegraphics[width=11em]{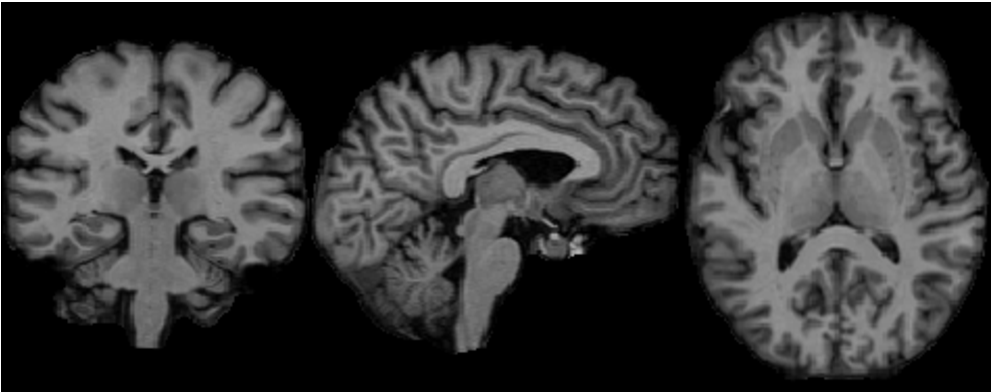}} & \begin{tabular}{cM{40mm}}         \scriptsize{Mean Significant Bits: 23.00} \\ \addimg{sub-0025598}{fs}{reg} \\ \scriptsize{Mean Significant Bits: 23.00} \\ \addimg{sub-0025598}{fs}{warp} \\ \end{tabular} & \begin{tabular}{cM{40mm}} \scriptsize{Mean Significant Bits: 19.42} \\ \addimg{sub-0025598}{sm}{reg} \\ \scriptsize{Mean Significant Bits: 18.63} \\ \addimg{sub-0025598}{sm}{warp} \\ \end{tabular} \\

    \multirow{5}{*}{2842950} & \multirow{2}{*}{\includegraphics[width=11em]{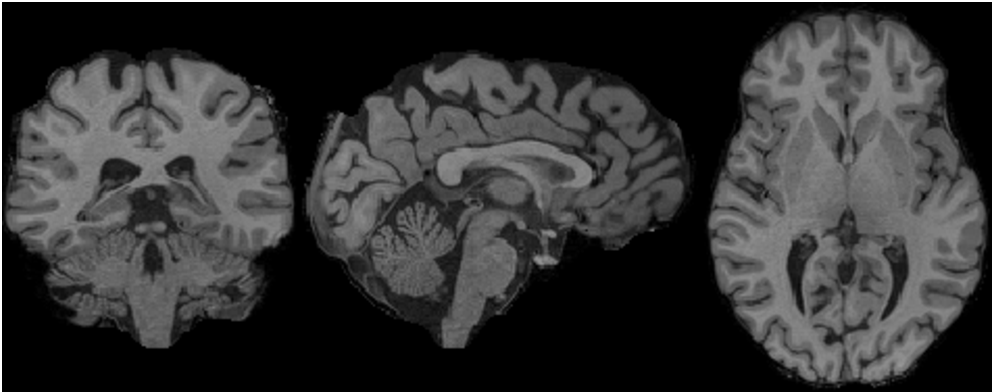}} & \begin{tabular}{cM{40mm}}         \scriptsize{Mean Significant Bits: 23.00} \\ \addimg{sub-2842950}{fs}{reg} \\ \scriptsize{Mean Significant Bits: 23.00} \\ \addimg{sub-2842950}{fs}{warp} \\ \end{tabular} & \begin{tabular}{cM{40mm}} \scriptsize{Mean Significant Bits: 19.49} \\ \addimg{sub-2842950}{sm}{reg} \\ \scriptsize{Mean Significant Bits: 18.59} \\ \addimg{sub-2842950}{sm}{warp} \\ \end{tabular} \\

  \end{tabular}

          \begin{tabular}{cM{40mm}M{40mm}M{40mm}}
    \multirow{5}{*}{0027094} & \multirow{2}{*}{\includegraphics[width=11em]{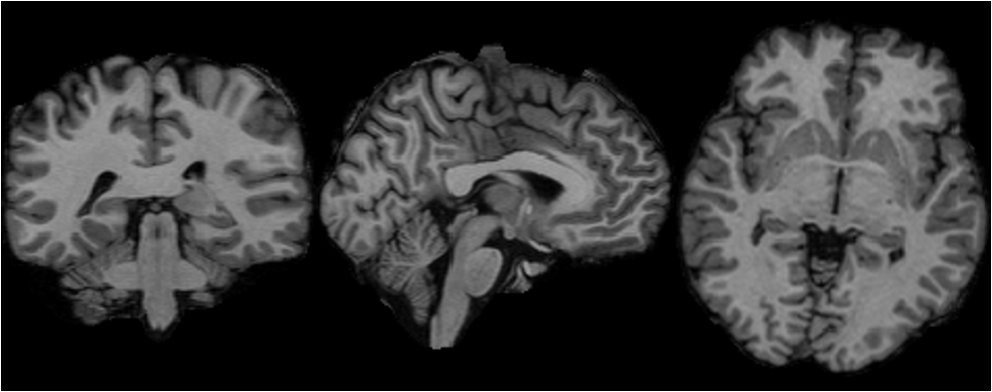}} & \begin{tabular}{cM{40mm}}         \scriptsize{Mean Significant Bits: 23.00} \\ \addimg{sub-0027094}{fs}{reg} \\ \scriptsize{Mean Significant Bits: 23.00} \\ \addimg{sub-0027094}{fs}{warp} \\ \end{tabular} & \begin{tabular}{cM{40mm}} \scriptsize{Mean Significant Bits: 19.40} \\ \addimg{sub-0027094}{sm}{reg} \\ \scriptsize{Mean Significant Bits: 18.60} \\ \addimg{sub-0027094}{sm}{warp} \\ \end{tabular} \\

    \multirow{5}{*}{0025406} & \multirow{2}{*}{\includegraphics[width=11em]{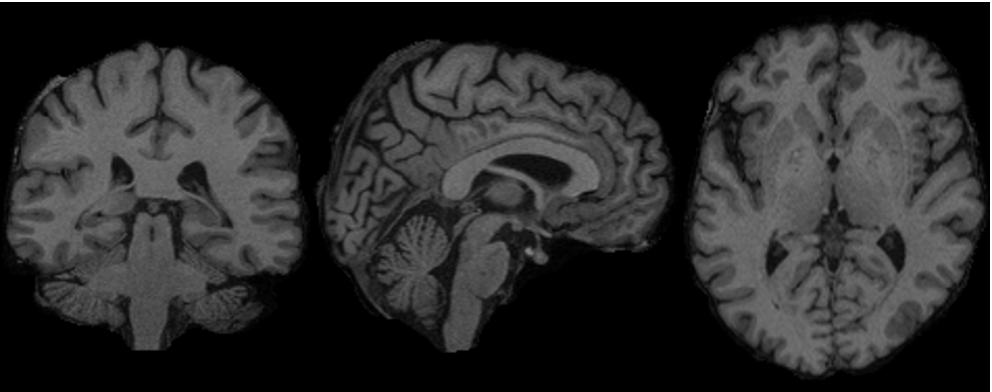}} & \begin{tabular}{cM{40mm}}         \scriptsize{Mean Significant Bits: 23.00} \\ \addimg{sub-0025406}{fs}{reg} \\ \scriptsize{Mean Significant Bits: 23.00} \\ \addimg{sub-0025406}{fs}{warp} \\ \end{tabular} & \begin{tabular}{cM{40mm}} \scriptsize{Mean Significant Bits: 19.44} \\ \addimg{sub-0025406}{sm}{reg} \\ \scriptsize{Mean Significant Bits: 18.46} \\ \addimg{sub-0025406}{sm}{warp} \\ \end{tabular} \\
    
    \multirow{5}{*}{0027236} & \multirow{2}{*}{\includegraphics[width=11em]{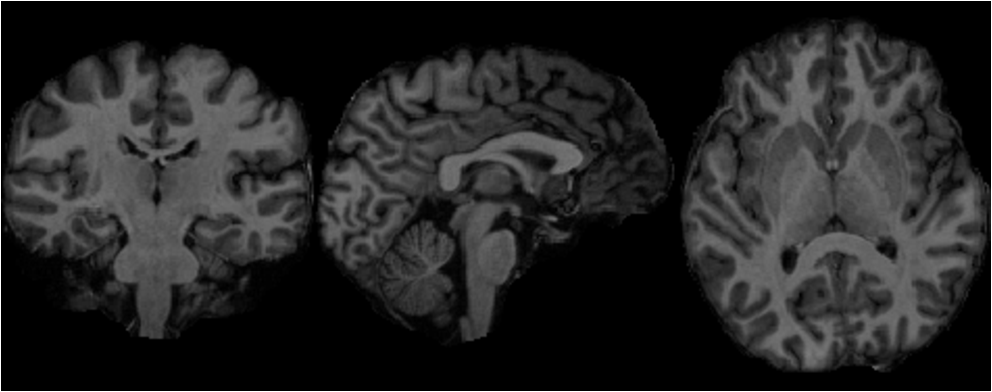}} & \begin{tabular}{cM{40mm}}         \scriptsize{Mean Significant Bits: 23.00} \\ \addimg{sub-0027236}{fs}{reg} \\ \scriptsize{Mean Significant Bits: 23.00} \\ \addimg{sub-0027236}{fs}{warp} \\ \end{tabular} & \begin{tabular}{cM{40mm}} \scriptsize{Mean Significant Bits: 19.21} \\ \addimg{sub-0027236}{sm}{reg} \\ \scriptsize{Mean Significant Bits: 18.59} \\ \addimg{sub-0027236}{sm}{warp} \\ \end{tabular} \\
    \bottomrule
  \end{tabular}
  \centering
    \includegraphics[width=0.6\textwidth]{figures3/colorbar.eps}
    \captionof{table}{Nonlinear registration uncertainty maps for images and their respective warp fields across FreeSurfer recon-all and SynthMorph. The displayed warp fields are averaged across the x, y and z axes. Due to required cropping of image dimensions, the registered brains do not have the same dimensions between SynthMorph and FreeSurfer, moreover the difference in methods produced differently sized warp fields.}

\section{Segmentation Entropy Maps for 32 Subjects}
\label{appendix:segmentation}

\begin{tabular}{cM{60mm}M{60mm}}
    \toprule
       Subject  & FreeSurfer recon-all & FastSurfer \\
    \midrule
    0026196 & \addpic{sub-0026196}{fs}{entropy} & \addpic{0026196}{fast}{entropy}\\
    0026044 & \addpic{sub-0026044}{fs}{entropy} & \addpic{0026044}{fast}{entropy}\\
    0027174 & \addpic{sub-0027174}{fs}{entropy} & \addpic{0027174}{fast}{entropy}\\
    0027268 & \addpic{sub-0027268}{fs}{entropy} & \addpic{0027268}{fast}{entropy}\\
    0027294 & \addpic{sub-0027294}{fs}{entropy} & \addpic{0027294}{fast}{entropy}\\
    0026017 & \addpic{sub-0026017}{fs}{entropy} & \addpic{0026017}{fast}{entropy}\\
    0025507 & \addpic{sub-0025507}{fs}{entropy} & \addpic{0025507}{fast}{entropy}\\
  \end{tabular}

\begin{tabular}{cM{60mm}M{60mm}}
    0025879 & \addpic{sub-0025879}{fs}{entropy} & \addpic{0025879}{fast}{entropy}\\
    0025350 & \addpic{sub-0025350}{fs}{entropy} & \addpic{0025350}{fast}{entropy}\\
    0025930 & \addpic{sub-0025930}{fs}{entropy} & \addpic{0025930}{fast}{entropy}\\
    0027214 & \addpic{sub-0027214}{fs}{entropy} & \addpic{0027214}{fast}{entropy}\\
    0025555 & \addpic{sub-0025555}{fs}{entropy} & \addpic{0025555}{fast}{entropy}\\
    0025436 & \addpic{sub-0025436}{fs}{entropy} & \addpic{0025436}{fast}{entropy}\\
    0025993 & \addpic{sub-0025993}{fs}{entropy} & \addpic{0025993}{fast}{entropy}\\
    0026175 & \addpic{sub-0026175}{fs}{entropy} & \addpic{0026175}{fast}{entropy}\\
  \end{tabular}

  \begin{tabular}{cM{60mm}M{60mm}}
    0025248 & \addpic{sub-0025248}{fs}{entropy} & \addpic{0025248}{fast}{entropy}\\
    0025531 & \addpic{sub-0025531}{fs}{entropy} & \addpic{0025531}{fast}{entropy}\\
    0025011 & \addpic{sub-0025011}{fs}{entropy} & \addpic{0025011}{fast}{entropy}\\
    0027012 & \addpic{sub-0027012}{fs}{entropy} & \addpic{0027012}{fast}{entropy}\\
    0027110 & \addpic{sub-0027110}{fs}{entropy} & \addpic{0027110}{fast}{entropy}\\
    0025362 & \addpic{sub-0025362}{fs}{entropy} & \addpic{0025362}{fast}{entropy}\\
    0025604 & \addpic{sub-0025604}{fs}{entropy} & \addpic{0025604}{fast}{entropy}\\
    0027191 & \addpic{sub-0027191}{fs}{entropy} & \addpic{0027191}{fast}{entropy}\\
  \end{tabular}
  
    \begin{tabular}{cM{60mm}M{60mm}}
    0003002 & \addpic{sub-0003002}{fs}{entropy} & \addpic{0003002}{fast}{entropy}\\
    0026055 & \addpic{sub-0026055}{fs}{entropy} & \addpic{0026055}{fast}{entropy}\\
    0025462 & \addpic{sub-0025462}{fs}{entropy} & \addpic{0025462}{fast}{entropy}\\
    0025631 & \addpic{sub-0025631}{fs}{entropy} & \addpic{0025631}{fast}{entropy}\\
    0025598 & \addpic{sub-0025598}{fs}{entropy} & \addpic{0025598}{fast}{entropy}\\
    2842950 & \addpic{sub-2842950}{fs}{entropy} & \addpic{2842950}{fast}{entropy}\\
    0027094 & \addpic{sub-0027094}{fs}{entropy} & \addpic{0027094}{fast}{entropy}\\
    0025406 & \addpic{sub-0025406}{fs}{entropy} & \addpic{0025406}{fast}{entropy}\\
    0027236 & \addpic{sub-0027236}{fs}{entropy} & \addpic{0027236}{fast}{entropy}\\
    \bottomrule
  \end{tabular}
  \centering
  \includegraphics[width=0.6\textwidth]{figures3/entropy_colorbar.eps}

\end{document}